\documentclass[useAMS,usenatbib]{mn2e}

\title[CASE: Dwarf nova in M55]
{Cluster AgeS Experiment (CASE): Detection of a dwarf nova in
the globular cluster M55}
\author[J. Kaluzny et. al.]
{J.~Kaluzny$^{1}$, P.~Pietrukowicz$^{1}$,
I.~B.~Thompson$^{2}$,\and
 W.~Krzeminski$^{3}$,
A.~Schwarzenberg-Czerny$^{1,4}$, W.~Pych$^{1}$, G.~Stachowski$^{1}$\\
  $^1$Nicolaus Copernicus Astronomical Center,
     ul. Bartycka 18, 00-716 Warsaw, Poland
     (jka,pietruk,alex,pych,gss@camk.edu.pl)\\
  $^2$Carnegie Institution of Washington, 813 Santa Barbara Street,
        Pasadena, CA 91101, USA
        (ian@ociw.edu)\\
  $^3$Las Campanas Observatory, Casilla 601, La Serena, Chile,
(wojtek@lco.cl)\\
  $^4$Astronomical Observatory, Adam
Mickiewicz University, ul. S{\l}oneczna 36, 61-286 Pozna\'n,
Poland\\
}

\date{Accepted .......
      Received .......;
      in original form ........}

\pagerange{\pageref{firstpage}--\pageref{lastpage}} \pubyear{2004}

\begin{document}

\maketitle

\label{firstpage}

\begin{abstract}
We report the detection of a dwarf nova (DN) in the core region of the
globular cluster  M55. Six outbursts were observed during 8
observing seasons spanning the period 1997-2004. The variable has
an X-ray counterpart detected on images taken with the $ROSAT$
telescope. Although we cannot offer proof of cluster membership,
one can see that both the position on the H-R diagram and the X-ray flux
are consistent with a bright DN at the cluster distance.
According to our outburst statistics, no more than one similar DN
could remain undetected in our field of view, centred at the
cluster core.

\end{abstract}
\
\begin{keywords}
stars: dwarf novae - novae, cataclysmic variables -- globular
clusters: individual: (M55, NGC 6809)
\end{keywords}

\section{Introduction} \label{s1}

It is well established on theoretical grounds that close
binary stars can have a significant impact on dynamical
evolution of globular clusters (Hut et al. 1992). At the same
time, the theory predicts that large numbers of white/red dwarf
binaries should form in the cores of globular clusters via
two-body tidal capture or three-body exchange capture (Fabian,
Pringle \& Rees 1975; Hut \& Paczy\'nski 1984). According to
expectations, some of these systems would be in orbits
sufficiently tight to become cataclysmic variables (CVs) at
some point in their evolution. For example, di~Stefano and
Rappaport (1994) estimate that there should be several thousand
CVs in the galactic globular clusters (GCs). Numerous ground-based
surveys for CVs in GCs yielded the identification of
surprisingly few objects (e.g. Shara et al. 1994). Besides the two
classical novae (Sawyer 1938; Hogg \& Wehlau 1964) observed in M14
and M80, a dwarf nova was detected in M5 (Oosterhoff 1941; Margon,
Downes \&Gunn 1981). Several dozen candidate CVs have been reported
over the last few years in observations collected with the {\it Chandra}
and {\it XMM-Newton} observatories (Grindlay et al. 2001; Gendre,
Barret \& Webb 2003) and with the {\it Hubble Space Telescope}
(e.g. Edmonds et al. 1999; Knigge et al. 2003). However, at present
there are just a few spectroscopically confirmed CVs in GCs
(Margon et al. 1981; Grindlay et al. 1995; Edmonds et al. 1999;
Knigge et al. 2003). Similarly, the list of objects with observed
dwarf nova type outburst is rather short: V101 in M5 (Shara,
Potter \& Moffat 1987), DN in NGC 6624 (Shara, Zurek \& Rich
1996), CV1 in M22 (Anderson, Cool \& King 2003), V1 in M15
(Charles et al. 2002), CV1 in M15  (Shara et al. 2004) and V2 in 47 Tuc
(Paresce \& de Marchi 1994).

We note parenthetically that at least 3 CVs have been identified so
far in open clusters, including one certain DN (Kaluzny et al.
1997; Gilliland et al. 1991). Recently
Mochejska et al. (2004) detected a very likely CV in the field of
the open cluster NGC~2158.
All 4 objects were discovered by
chance and one may expect the detection of more CVs, even in close
open clusters, if a systematic survey is undertaken.

We have started a systematic search for erupting dwarf novae in
GCs by taking advantage of the rich photometric data base collected
by the CASE collaboration. The CASE project is focused on
the determination of ages and distances to nearby GCs using
observations of detached eclipsing binaries (Kaluzny et al. 2005).
To identify suitable binaries we conducted an extensive
photometric survey of a dozen GCs. The data obtained during the
survey can be used to search for various types of variable objects
and, in particular, for optical transients in the fields of the clusters
being monitored. This paper presents results obtained for M55,
which was the first cluster searched by us for the presence of 
possible dwarf novae.

%CVs in open clusters.

%The Bright Nova of 1860 in the Globular Cluster Messier 80. Sawyer 1938
% Hogg \& Wehlau 1964 nova in M14
%cluster was discovered in M-14 by Amelia Wehlau of the University of
%Western Ontario. The nova actually brightened in 1938 and was found at
%the later date from plates taken by Helen Sawyer Hogg between 1932 and
%1963, using the 74-inch reflector of the David Dunlap Observatory and
%the 72-inch of the Dominion Astrophysical Observatory. 255 photos were
%obtained on 124 nights during this period and the nova appears on only
%eight of them, these being taken on June 21-28th, 1938. The only other
%nova found in a globular cluster was T Scorpii, discovered in 1860 in
%M-80.

\section{Observations}\label{s2}

The CASE project is conducted at Las Campanas Observatory. For the
survey we used the 1.0-m Swope telescope, equipped with a
$2048\times 3150$ pixel SITE3 CCD camera\footnote{The actual size
of the CCD is $2048\times 4096$ pixels, but rows above 3150 are
unusable for photometry.}. With a scale of 0.435 arcsec/pixel, the
usual field of view is about $14.8\times 23$~arcmin$^{2}$. However, a
large fraction of images for the M55 cluster were taken with a
smaller subraster covering a field of $14.8\times 11.6$~arcmin$^{2}$
(longer axis aligned in the E-W direction). In the present search for
CVs all of the analysed frames were trimmed to this smaller size. 
The cluster core was located approximately at the centre of the
subraster field.

The cluster M55 was monitored during 8 observing seasons spanning
years 1997-2004. A total of 3795 images were taken through the $V$
filter, with exposure times ranging from 100s to 480s, while
exposure times for a further 313 images taken in the $B$ filter ranged from 100s
to 420s. The number of $V$-band frames taken per night ranged from
a few to 90. The median seeing
was 1.44  and 1.51 arcsec for the $V$ and $B$ bands,
respectively.
The cluster was also observed during the 1997 and 2003  seasons
with  the TEK5 direct CCD camera attached to the 2.5-m du~Pont
telescope. The field of view was $8.8\times 8.8$~arcmin$^{2}$ at
a scale of 0.259 arcsec/pixel. The time series photometry
through $BV$ filters was obtained on a total of 6 nights in
May-June 1997 and on 3 nights in May 2003. In addition, exposures
with $UBVI_{\rm C}$ filters  were collected on the
nights of 1997 June 2 and 2003 May 11. The journal of
observations used to extract $UBVI_{\rm C}$ photometry discussed
in Sec. \ref{f4} is listed in Table \ref{t1}.

\begin{table}
 \centering
 \begin{minipage}{200mm}
  \caption{$UBVI_{\rm C}$ observations of M55.\label{t1}}
{\small
  \begin{tabular}{lcrr}
\hline
Date & Filter & Exposure & Seeing \\
     &        &  (s)     & (arcsec)\\
\hline
1997 June 2 & V & 5$\times$ 45  & 0.75 \\
1997 June 2 & B & 6$\times$ 65  & 0.82 \\
1997 June 2 & V & 7$\times$ 35  & 0.73 \\
1997 June 2 & I & 3$\times$ 25  & 0.61 \\
1997 June 2 & U & 5$\times$ 360 & 0.81 \\
1997 June 2 & V & 8$\times$ 35  & 0.81 \\
2003 May 11 & V & 2 $\times$ 30  & 0.99   \\
2003 May 11 & I & 2 $\times$ 20  & 0.83   \\
2003 May 11 & U & 2 $\times$ 120 & 1.20   \\
2003 May 11 & B & 2 $\times$ 50  & 1.13   \\
2003 May 11 & V & 2 $\times$ 35  & 1.25   \\
\hline
\end{tabular}}
\end{minipage}
\end{table}

\section{Search for Dwarf Novae}\label{s3}
The search for possible DNe in M55 was conducted on the $V$ filter
images collected with the Swope telescope.  We used a slightly
modified version of the \textsc{isis-2.1} image subtraction package (Alard
\& Lupton 1998; Alard 2000) to detect the variable objects and to
extract their photometry. Our procedure followed that described in
detail by Mochejska et al. (2002). A reference image was
constructed by combining 19 individual frames with $T_{exp}=120{\rm s}$
taken during dark time on the night of 2001 July 12/13. The
seeing for the resultant reference image was $FWHM=1.00$~arcsec
and the limiting magnitude corresponding to a $3\sigma$ 
detection level was $V\approx 23.0$.\footnote{This
limiting magnitude only applies to the least crowded part of the
analysed field.} Subsequently we selected the nights for which at
least two $V$ images with seeing better than 1.6 arcsec were
available. There were 145 such nights and for 113
of them it was possible to select at least  5 images fulfilling
the above condition. The data sets consisting of 2-5 images were
then combined to form an average image for each of the 145
nights. Use of the combined frames to search for erupting
variables is advantageous not only because of the S/N issue but
also because the combined images are free from defects caused by
cosmic rays. The combined images were remapped to the reference
image coordinate system and subtracted from the
point spread function (PSF) convolved reference image using programs
from the \textsc{isis} package. The resultant frames were searched
with \textsc{daophot} (Stetson 1987) for presence of any subtraction
residuals with stellar PSF profiles. We omitted from the search regions 
which corresponded to the location of saturated stars or to
known variables (Clement et al. 2001; Pych et al. 2001; a more
extended list based on CASE results will be published elsewhere).
In addition, to avoid too many false alarms
we set a high detection threshold, at a total residual flux
equivalent to that of a constant star with $V=20.5$. In other
words, such a star would be marginally detected if it doubled its
flux. The apparent distance modulus of M55 is 
$(m-M)_{V}=13.87$ (Harris 1996). At this distance our variability limit in terms of
excess flux corresponds to the constant flux produced by a star
of $M_{V}=6.6$.

Our analysis yielded the identification of just one certain erupting
object, which we shall call CV1. Its equatorial
coordinates are: $\alpha_{J2000}=19^{h}40^{m}08.^{s}59$,
$\delta_{J2000}=-30^{\degr} 58^{\arcmin} 51.^{\arcsec}1$. The
external accuracy of these coordinates is about 1.0\arcsec.

For further analysis, the $BV$ light curves of the variable were
extracted with \textsc{isis}, using individual images rather than the
combined frames. In the case of the $B$-band photometry, the
reference image was constructed by combining the 15 best
seeing exposures from the 1998 season. The reference
image has $FWHM=1.1$~arcsec and the limiting magnitude is
$B\approx 23.2$ The instrumental photometry was transformed to the
standard $BV$ system using a set of secondary standards present in
the M55 field (Stetson 2000). Specifically we used a total of 49
standards with $0.07<B-V<1.04$.

\section{Properties of CV1}\label{s4}
\begin{figure}
\begin{center}
\vspace{15.9cm}
\includegraphics{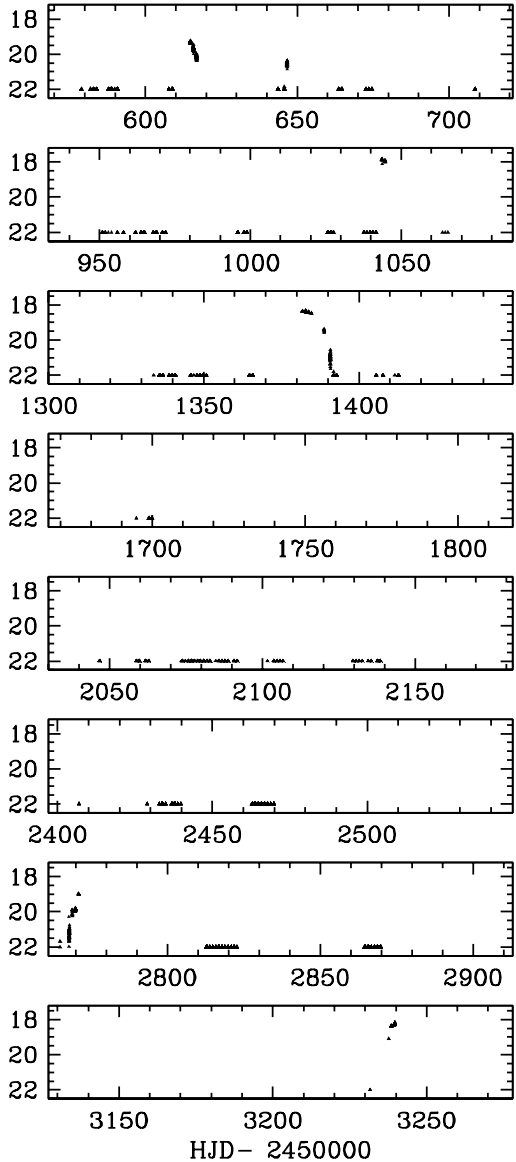}
\end{center}
\caption{\label{f1} $V$-band light curve of CV1 for observing
seasons from 1997 (top) to 2004 (bottom). Each panel covers the
period May-September for a given year.}
\end{figure}

\begin{figure}
\begin{center}
\vspace{7.0cm}
\includegraphics{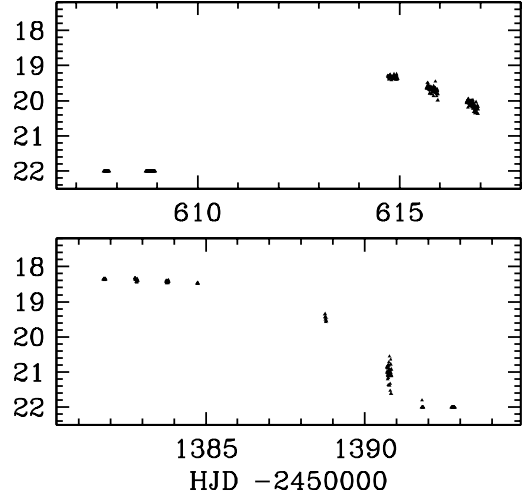}
\end{center}
\caption{\label{f2} $V$-band light curve of CV1 covering outburst
observed in 1997 (top) and 1999 (bottom) seasons. }
\end{figure}

Our observations of the variable CV1 recorded 6 outbursts. The
combined light curve in the $V$-band is presented in Fig.~\ref{f1},
while Fig.~\ref{f2} shows two selected outburst light curves. On
individual out-of-outburst images collected with the 1.0-m Swope
telescope, the CV1 variable is, in fact, below or close to the
detection limit. In Figs. \ref{f1} and \ref{f2} the data points
corresponding to the out-of-outburst images are plotted assuming
$V=22$. 
\begin{figure}
\begin{center}
\vspace{5.0cm}
\includegraphics{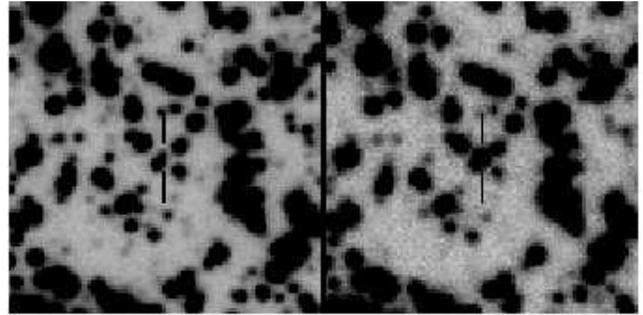}
\end{center}
\caption{\label{f3} Finding charts for CV1 showing the variable in
quiescence (left) and in outburst (right). North is up, and east
is to the left. The field of view is $26~\arcsec$ on a side.}
\end{figure}

The variable in its low state is still detectable on the
reference images, but even then the exact measurement of its
magnitude is difficult because of crowding problems and relatively
poor sampling of the PSF. Photometry of the variable in quiescence
is much easier on the images collected with the du~Pont telescope.
We have used the data listed in Table 1 to measure magnitudes and
colours of CV1 in its low state (1997 June 2) and on the rising
branch of an outburst (2003 May 11).\footnote{Observations
collected with the du~Pont telescope show the variable at $V\approx
20.7$ on 2003 May 4 and at $V\approx 19.9$ on 2003 May 10.}
Figure \ref{f3} shows images of the field of the variable for
these two nights. The close companion visible south-east of CV1 is
located at an angular distance of 0.94~arcsec from the variable
and has $V=20.15$ and $B-V=0.465$. The following magnitude and
colours of the variable were derived from the images taken on June
2, 1997: $V=21.88\pm 0.06$, $B-V=0.63\pm 0.08$, $U-B=-0.83\pm
0.09$ and $V-I=1.18\pm 0.09$. For the night of May 11, 2003 we
obtained: $V=18.98\pm 0.02$, $B-V=0.13\pm 0.02$, $U-B=-0.66\pm
0.03$ and $V-I=0.26\pm 0.03$. The instrumental photometry was
transformed to the standard system using observations of
several Landolt (1992) fields. The 1$\sigma$ uncertainties quoted
only reflect internal errors and do not include the uncertainties of
the zero points of the photometry. To check the external errors
we have compared our photometry against standard stars visible
in the cluster field (Stetson 2000). For 28 stars found in common the
average differences for $V$, $B-V$ and $V-I$ amount to 0.022,
0.022 and 0.033 mag, respectively. No such comparison was possible
for the $U-B$ colour as no Stetson photometry is available for the
$U$ band. The data obtained  with the Swope telescope give the
median colour of the variable during outburst as $B-V=0.12$.

\begin{figure*}
\begin{center}
\vspace{8.3cm}
\includegraphics{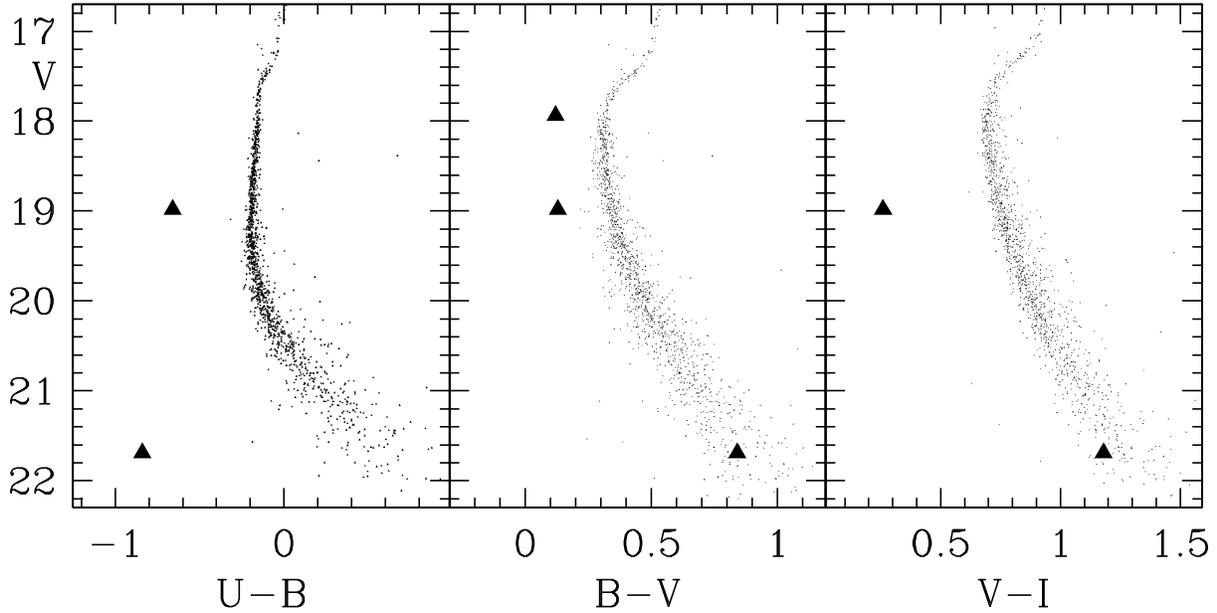}
\end{center}
\caption{\label{f4} Location of CV1 on colour-magnitude diagrams
of M55, in quiescence ($V=21.8$; 1997 June 2), on the rise to
maximum ($V=19.0$; 2003 May 11) and at maximum  ($V=17.9$; 1998 Aug
19).}
\end{figure*}
\begin{figure*}
\begin{center}
\vspace{8.0cm}
\includegraphics{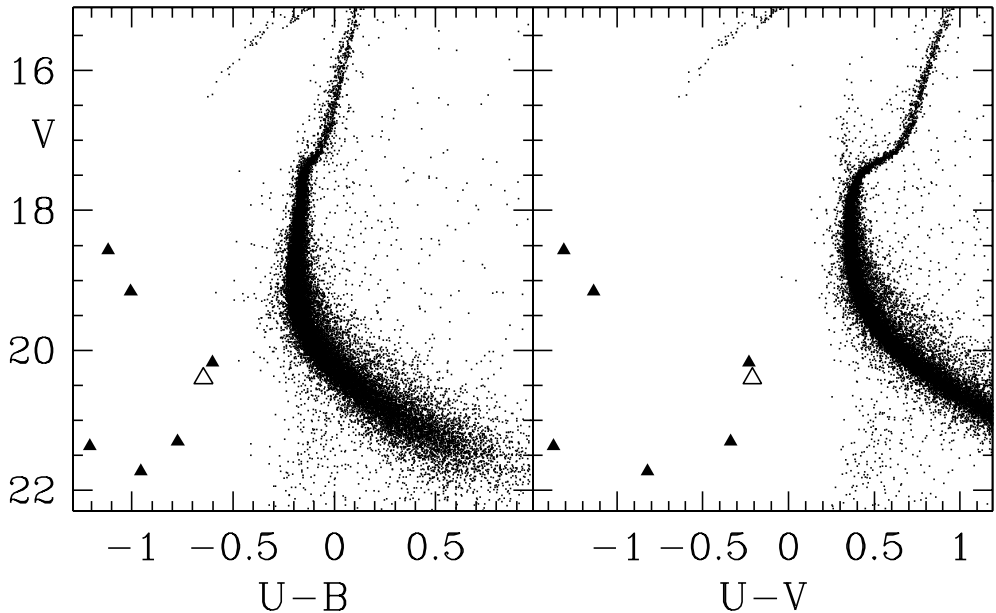}
\end{center}
\caption{\label{f5} Colour-magnitude diagrams of M55. Stars with $U-B<-0.6$
are marked with triangles. The open symbol corresponds to the variable M55-B1.}
\end{figure*}

The nature of the variability of CV1 along with its observed colours
indicates that it is a cataclysmic variable of dwarf nova type.
According to Warner (1976), most DNe at maximum outburst have
unreddened colours in the range $(B-V)_{0}=0.0\pm 0.10$ and
$(U-B)_{0}=-0.80\pm 0.15$. It appears that the colours displayed by CV1
during outburst fall within the above ranges, especially if we
take into account the reddening of M55, which amounts to  $E(B-V)=0.08$
(Harris 1996).

In Fig. \ref{f4}, we show the location of CV1 in three different
luminosity stages on the colour-magnitude diagram of the cluster.
It is worth noting that in the low state the variable is located
close to the cluster main sequence on the $V/B-V$ and
$V/V-I$ planes, although it is still very blue in the $U-B$ colour.
There are several examples of DNe from GCs and open clusters which
show relatively red optical colours in quiescence (Kaluzny \&
Thompson 2003; Kaluzny et al. 1997; Anderson, Cool \& King 2003).
Hence, CV1 is hardly exceptional in that respect (see also Bruch \& Engel 1994).
Out of a total of 193 nights, CV1 was seen in outburst on 23 nights,
yielding an average duty cycle of $\rho= 23/193 = 0.119\pm0.025$. Due to
telescope scheduling and weather, the available nights were not distributed
randomly and tend to clump, however, both the numerator and denominator
should be affected in a similar way without any systematic effect
on the duty cycle. The best-covered outburst lasted $t \approx 10$ days.
Each outburst was on average covered by 4, not necessarily consecutive, nights.
Hence we estimate that, to within 30 percent accuracy, the
above duration $t$ is typical for all outbursts. This yields the
average outburst cycle length $T = t/\rho \approx  84$ days with a
similar 30 percent error. These results might be used to deduce
the average accretion rate and evolution time of the binary, a
subject outside our present scope.

We are not in a position to determine from our observations the total
number of close binary stars in the cluster core, an important
number for its dynamical evolution. However, we can try to
estimate the maximum number of DNe in the cluster which would still be consistent
with our observations. Let us assume that in our field there is
another DN with properties identical to CV1. For as small a duty
cycle as $\rho=0.12$, its outburst should, to a good
approximation, obey the Poisson distribution $P(r\leq k;\lambda)$, with the
average number of outbursts $\lambda = 6$ from CV1. However, we
did not observe any outbursts from a star other than CV1 $(k=0)$, an
occurance of probability $P(r=0; 6) = 0.0025$ or $2.4\sigma$
significance. As a conservative estimate, we assume
that a single outburst per star could be misinterpreted as an
artefact and overlooked; thus we obtain a probability $P(r\leq 1; 6)^2
= 0.0003$ for the presence of as many as 2
undetected DNe in our field. Therefore, the hypothesis that M55 contains as many as 3 DNe
including CV1 has to be rejected at the $3.7\sigma$ confidence level.
Furthermore, the peak outburst magnitude of CV1 is around $V=18.0$. However, in our
photometry, outbursts as faint as $V\approx19.8$ mag would stand out. Hence
we can be confident of our conclusion on the lack of 3 DNe in the cluster with
outburst magnitude $M_V<5.9$.

The available data do not allow us to establish with confidence the
cluster membership status of CV1. However, we note that the range
of observed luminosities of the variable, $18<V<21.8$, is
consistent with the assumption that it is a dwarf nova belonging to
M55. The cluster has an apparent distance modulus $(m-M)_{V}=
13.87$ (Harris 1996) which yields a variability range in absolute
magnitudes of $4.3<M_{V}<7.9$ for CV1 under the assumption of cluster
membership. Such a range would be normal for a bright,
non-magnetic DN.

The X-ray observations conducted with the $ROSAT$ PSPC detector
led to the detection of 18 sources in the field of M55 (Johnston,
Verbunt \& Hasinger 1996) of which one, namely object \#9, is a very
likely counterpart of CV1. For that object the  {\it ROSAT Source
Catalog of Pointed Observations with the HRI} (ROSAT Team, 2000)
gives $\alpha_{2000}=19^{h} 40^{m} 08.4^{s}$ and $\delta_{2000}
=-30\degr 58\arcmin 52\arcsec$ with a positional error of 2~arcsec.
The optical coordinates of CV1 listed above fall within the error
circle of the $ROSAT$ source M55-\#9. The X-ray to optical
luminosity ratio $L_x/L_o \approx 0.3$ would be higher than
average for DN but consistent with that for SS Cyg, assuming CV1
was in quiescence during $ROSAT$ observations.

%1RXH J194008.4-3035852

\section{Blue stars in the cluster field}\label{s5}
\begin{figure}
\begin{center}
\vspace{5.0cm}
\includegraphics{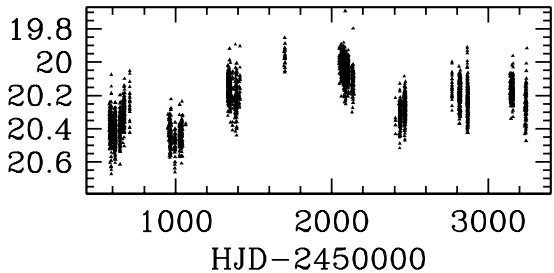}
\end{center}
\caption{\label{f6} $V$-band light curve of blue star M55-B1.}
\end{figure}

There are several classes of CVs which do not show any pronounced 
outbursts on a time-scale of years or even decades. For example, old
novae or AM CVn stars typically show variability on the level of a few tenths
of a magnitude in the optical domain and on time-scales ranging from 
seconds to years. However, most CVs show very blue
$U-B$ and $U-V$ colours (Bruch \& Engel  1994). Hence, one may search for candidate 
CVs in globular clusters by looking for  blue stars located below 
the horizontal branch on the $U-B/V$ or $U-V/V$ colour-magnitude diagram.
We have constructed such diagrams for M55 using the data collected with the 
du~Pont telescope on the night of 1997 June 2. The photometry was extracted
from the following combined images: $U$~$5\times 360s$, $B$~$6\times 65s$,
$V$~$40\times 35s$, $V$~$5\times 10s$. The resultant colour-magnitude diagrams
are shown in Fig. \ref{f5}. We considered seven blue objects with measured $U-B<-0.6$
to be candidate CVs (CV1 was dropped from the list). Not one of the X-ray sources detected
in the M55 field by the $ROSAT$ PSPS detector (ROSAT Team, 2000) is located within 
10 arcsec of any of these blue stars. Examination of time series photometry based on 
the data from the Swope telescope allowed us to detect variability for only one of candidates.
This object, which we call M55-B1, is seen at $V=20.40$ and $U-B=-0.65$ in 
Fig. 5 and its equatorial  coordinates are: 
$\alpha_{J2000}=19^{h}39^{m}49.^{s}81$,
$\delta_{J2000}=-30^{\deg} 53^{\arcmin} 19.^{\arcsec}6$.
It exhibits season to season
changes of the mean $V$ luminosity on a level of a few tenths of a magnitude. The light
curve for the period 1997-2004 is shown in Fig. \ref{f6}.   
In order to  search for possible short term variability of blue star candidates for CVs
we have examined  their time series  photometry extracted with \textsc{isis} 2.1 (Alard 2000)
from the data collected with the du~Pont telescope in 1997. A total of 592
$V$ images with exposure times ranging from 30~s to 120~s were  taken during the 
period 1997 May 31 -- 1997 June 7. 
Time series photometry was extracted from individual images as well as  
from 64  combined images formed by averaging 5-10 consecutive frames. 
We failed to detect any significant variability for any of the blue candidates marked
in Fig. \ref{f5}. In particular, the light curve of variable M55-B1 
based on combined images
was flat, with $<V>=20.40$ and $rms=0.014$ mag. We conclude that none of the selected blue
stars is a likely candidate for a CV. The blue colours and long time-scale low-amplitude 
variability of M55-B1 make it a likely  candidate for a quasar. \\
We conclude this section by noting that a large fraction of stars with $V>18.5$  visible 
to the blue or below the main-sequence of M55 in Fig. \ref{f5} are members of  the Sagittarius 
dwarf galaxy which is located in the background of the cluster (Mateo et al. 1996). It is 
feasible that some of blue stars considered above belong to the extended blue horizontal
branch of the Sagittarius dwarf.

\section{Summary}\label{s6}

We report the results of a search for DN outbursts in an extensive
photometric survey of the globular cluster M55 obtained within the
scope of the CASE collaboration. Using a total of 3795 $V$-band images
taken over 8 seasons centred on the cluster core we were only able to
detect 6 outbursts, all from the newly discovered DN star
CV1. As we relied on a median combination of several images and on
subtraction of the PSF matched templates using the \textsc{isis} technique,
our survey is quite deep and sensitive down to the very crowded
cluster core. Our outburst statistics are consistent with the
absence of any further undetected DNe similar to CV1 in the
investigated field, and we reject at the $3.7\sigma$ confidence
level the hypothesis that there are 2 additional undetected DNe with
outbursts similar to CV1 or fainter, down to $M_V\approx 5.9$. While most
bright field DNe located at the distance of M55 would be detected
in our survey, we caution that DNe are in general heterogeneous
class of objects and cluster CVs differ in metallicity, so that
some rather exotic faint DNe and/or with rare outbursts could
perhaps have escaped our scrutiny. However, the deficit of DNe in 
globular clusters appears to be real. Generally, any dynamical evolutionary
scenario of cluster cores would thus have to address the issue of the slow
creation of DNe and/or destruction of the existing ones.

The outbursts of CV1 last about 10 days and recur every 85 days, on
average. Its position appears to coincide with a $ROSAT$ point
source. Although we do not offer proof of the cluster membership of CV1,
its characteristics -- position in quiescence on the
colour-magnitude diagram and X-ray flux -- are entirely consistent
with a fairly bright dwarf nova $M_{V}(min) = 7.9$ located at the
distance of M55.

In addition, we searched our deep multicolour photometry obtained
with the du~Pont telescope for the presence of any blue objects. Apart
from CV1, we discovered another blue object showing variability between
seasons with no evidence of short time scale  variability. Its properties suggest
that it is quite likely to be a background quasar. 
%We caution that completeness of this part of our survey is
%difficult to establish.

\section*{Acknowledgments}
PP was supported by the grant 1~P03D 024 26 from the 
State Committee for Scientific Research, Poland. JK \& GS were
supported by the grant 1~P03D 024 26 from the
Ministry of Scientific Research and Informational Technology, Poland.

\label{lastpage}

\end{document}